\documentclass[11pt]{article}
\usepackage{a4}
\usepackage{amsmath}
\usepackage{amsfonts}
\usepackage{latexsym}
\usepackage{epsfig}
\newcommand{\be}{\begin{equation}}
\newcommand{\ee}{\end{equation}}
\newcommand{\bea}{\begin{eqnarray}}
\newcommand{\eea}{\end{eqnarray}}

\renewcommand{\d}{\mathrm{d}}

\DeclareMathSymbol{\mg}{\mathrel}{symbols}{"1D}

\newcommand{\ga}{\alpha}

\newcommand{\gf}{\phi}

\newcommand{\gx}{\xi}
\newcommand{\gm}{\mu}
\newcommand{\gn}{\nu}

\newcommand{\gl}{\lambda}
\newcommand{\gr}{\rho}

\newcommand{\gz}{\zeta}
\newcommand{\gp}{\pi}
\newcommand{\gps}{\psi}

\newcommand{\gG}{\Gamma}

\newcommand{\gS}{\Sigma}

\newcommand{\cP}{{\cal P}}

\newcommand{\cR}{{\cal R}}

\newcommand{\ra}{\rightarrow}

\renewcommand{\Im}{\text{Im}}

\newcommand{\dsp}{\displaystyle}

\newcommand{\labl}[1]{\label{#1}}
\newcommand{\half}{\frac 12 }
\newcommand{\shalf}{{\scriptstyle \half}}

\newcommand{\beq}{\begin{equation}}
\newcommand{\eeq}{\end{equation}}
\newcommand{\barr}{\begin{array}}
\newcommand{\earr}{\end{array}}
\newcommand{\equ}[1]{\begin{gather} #1 \end{gather}}

\newcommand{\mtrx}[1]{\begin{matrix} #1 \end{matrix}}

\newcounter{oldcounter}

\newcommand{\Intr}{\mathbb{Z}}

\begin{document}

\begin{flushright} 
hep-th/0111288
\end{flushright} 
\vskip 2 cm
\begin{center}
{\Large {\bf 
Kaluza-Klein towers on orbifolds: divergences and anomalies
}}
\\[0pt]
\bigskip {\large
{\bf Stefan Groot Nibbelink\footnote{
{{ {\ {\ {\ E-mail: nibblink@th.physik.uni-bonn.de}}}}}} }
\bigskip }\\[0pt]
\vspace{0.23cm}
{\it Physikalisches Institut der Universitat Bonn,} \\
{\it Nussallee 12, 53115 Bonn, Germany.}\\
\bigskip
\vspace{3.4cm} Abstract
\end{center}
{\small
The ultra-violet behavior of Kaluza-Klein theories on a one dimensional
orbifold is discussed. An extension of dimensional regularization that
can be applied to a compact dimension is presented. 
Using this, the FI-tadpole is calculated in the effective KK theory
resulting from compactifying supersymmetric theories in 5 dimensions. 
\\[1ex]
Talk presented at :
\begin{center}
\begin{minipage}{5cm}
    COSMO-01 \\
    Rovaniemi, Finland, \\
    August 29 -- September 4, 2001
\end{minipage}
\end{center}
}
\newpage

\section{Introduction}

Models with 5 dimensional global supersymmetry compactified on 
orbifolds may be good candidates for extensions of the
standard model and may have interesting cosmological applications. 
The supersymmetry may give rise to many impressive
ultra-violet properties while the orbifold compactification can produce to 
phenomenologically interesting particle spectra. 
One such setup proposed by Barbieri, Hall, Nomura (BHN)
\cite{Barbieri} has particular remarkable properties. 
This model has the low energy spectrum identical to the standard model
including a single massles Higgs, obtained by compactifying a
supersymmetric theory with vector and hyper multiplets on the orbifold
$S^1/\Intr_2 \!\times\! \Intr_2'$ in 5 dimensions. 
In the following table the Kaluza-Klein spectrum of the BHN model is
presented. 
\begin{center}
\scalebox{.75}{\mbox{
\includegraphics*[40mm,210mm][165mm,280mm]{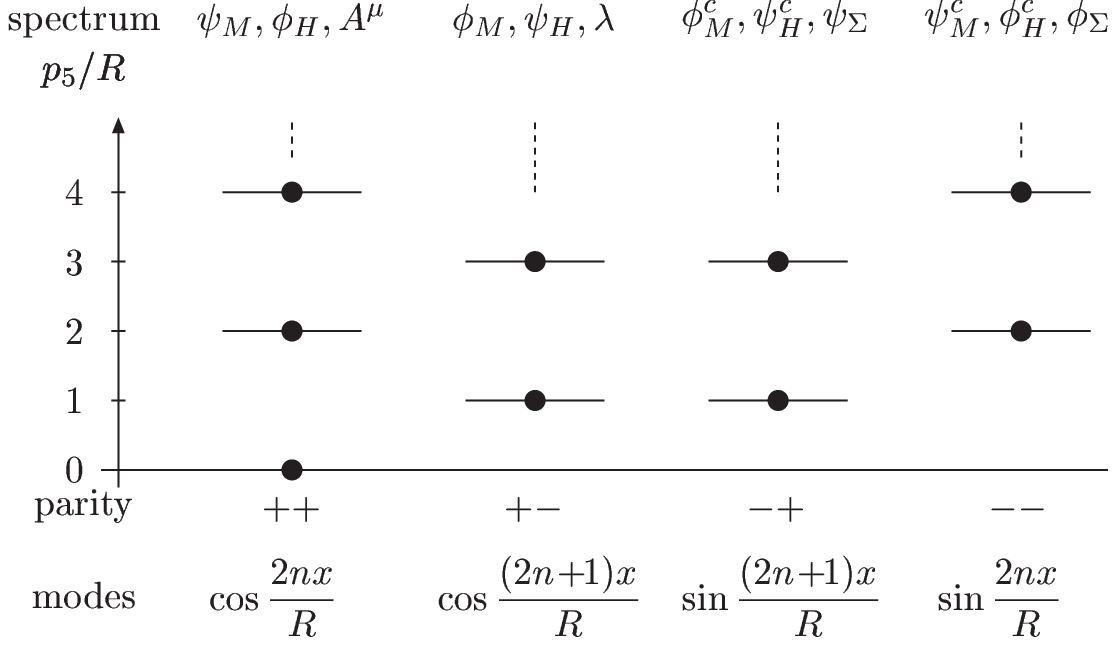}
}}
\end{center}
The parities dictate in which mode functions a given field has to be
expanded. The field theory consists of a complex Higgs scalar $\gf_H$,
its Higgsino $\gps_H$, the standard model fermions $\gps_M$, 
their mirrors $\gps_M^c$, 
the standard model gauge fields $A_\gm$, the two gauginos $\gl,
\gps_\gS$ andthe sfermions $\gf_M, \gf_M^c$. 
The 5th component of the gauge field $A_5$ in 5 dimension
and an additional scalar reside in $\gf_\gS$. Finally the 
superscript $^c$ denotes
independent charge conjugate states. These states form vector and hyper
multiplets of the original supersymmetric theory. 

One can raise various questions about the physical properties of such
models with towers of KK states:
Can we make sense of {infinitely} many fields (at the quantum level)?
Can low energy {anomalies} arise?
What happens in the {limit of very large radius $R \ra \infty$}? 
Is a theory with KK excitations ``better'' behaved than the
{low-energy} its theory? 

It was claimed in the recent literature that these
types of orbifold models have a extremely mild ultra-violet
behavior \cite{finite_results}:
the effective action was claimed to be finite at one loop or even to
all orders in perturbation theory. Others \cite{Ghilencea}
raised objections to such strong (unmotivated) claims. In ref.\
\cite{SHPD} it was shown that the BHN model has a quadratic divergence
due to a Fayet-Iliopoulos (FI) term and therefore its UV behavior is 
similar to that of the ordinary standard model. 

In this proceedings we report on this work using technique of 
dimensional regularization of a compact dimension introduced in ref.\
\cite{sgn}. After this method is described, it is used to calculate
the quadratically divergent FI term and to confirm some naive intuition
concerning anomalies in low-energy effective theories. The talk ends
with a short conclusion.

\section{Dimensional regularization of a compact dimension} 
\labl{sumasint}

Dimensional regularization in $4$ uncompact dimensions 
is a very powerful and convenient regularization scheme \cite{collins}
because it relies on properties of complex functions.
It is universal in the sense that it can be applied to an arbitrary 
loop calculation. 
It respects all symmetries of the classical theory, except when this symmetry 
develops an anomaly at the quantum level.

Dimensional regularization has been used before in the connection 
with compact manifolds. For example, in ref.\ 
\cite{Candelas,Sochichiu,DiClemente} 
it was combined with $\gz$-function regularization 
(see ref.\ \cite{Elizaldeetal} for a general review of this method) 
for the compact dimension. 
The crucial difference with the method introduced in \cite{sgn} is
that there are two independent regulators $D_4$ and $D_5$ 
for the 4 dimensional integration and the additional summation 
of KK momenta. 

It is not possible to directly apply the standard dimensional
regularization techniques to a field theory defined on a compact
dimension. We describe a procedure \cite{sgn} 
how this can be done in two steps: 
1.\ rewrite the sum over KK excitations as a
complex contour integral, 2.\ modify this integral by inserting a
regulator function. For concreteness we consider the infinite sum 
\equ{
\sum_{n \geq 0}  \frac 1{p_4^2 + {(2n)^2} {\scriptstyle R^{-2}}} = 
\half \sum_{{n \in \Intr} } \frac 1{p_4^2 + {(2n)^2} 
{\scriptstyle R^{-2}}} + {\half \frac 1{p_4^2}}. 
}
Here we have used that the summation is symmetric under $n \ra -n$. 
This can be represented as an integral along a (clockwise) contour
$\rightleftharpoons$ around the real axis \cite{Arkani,Mirabelli}, 
given in the picture below,  
\equ{ 
    \int_{{\rightleftharpoons}}  \frac {{-\d p_5}}{2\gp i} 
 \frac {\cP^{++}(p_5)}{p_4^2 + p_5^2} 
=  \int_{{\ominus}}  \frac {\d p_5}{2\gp i} 
\frac {\cP^{++} (p_5)}{p_4^2 + p_5^2 +m^2},
\quad 
{\cP^{\pm\pm}} = {\half}  
{\left(
{\pm\frac {1}{p_5}} + \frac {\shalf \gp R}{\tan \shalf \gp R p_5} 
\right)}, 
\labl{polefunS22}
}
where we have introduce the ``pole functions''  $\cP^{\pm\pm}$. 
(In the computation of the Fayet-Iliopoulos term we use $\cP^{--}$ as well.)
By noticing that the function 
$1/(p_4^2 +{(2n)^2} {\scriptstyle R^{-2}})$ does not have a pole at
infinity, it follows that this integral can be rewritten as contour
integral $\ominus$ over the upper and lower half plane with opposite
orientation  (anti-clockwise) to the $\rightleftharpoons$ contour. 
(We have introduced 
an IR regulator mass $m$, which is needed to turn the sum into 
a contour integral by complex function analysis \cite{sgn}.)
The figure below gives a schematic picture of this situation in the 
complex $p_5$-plane: 
\begin{center}
\epsfig{file=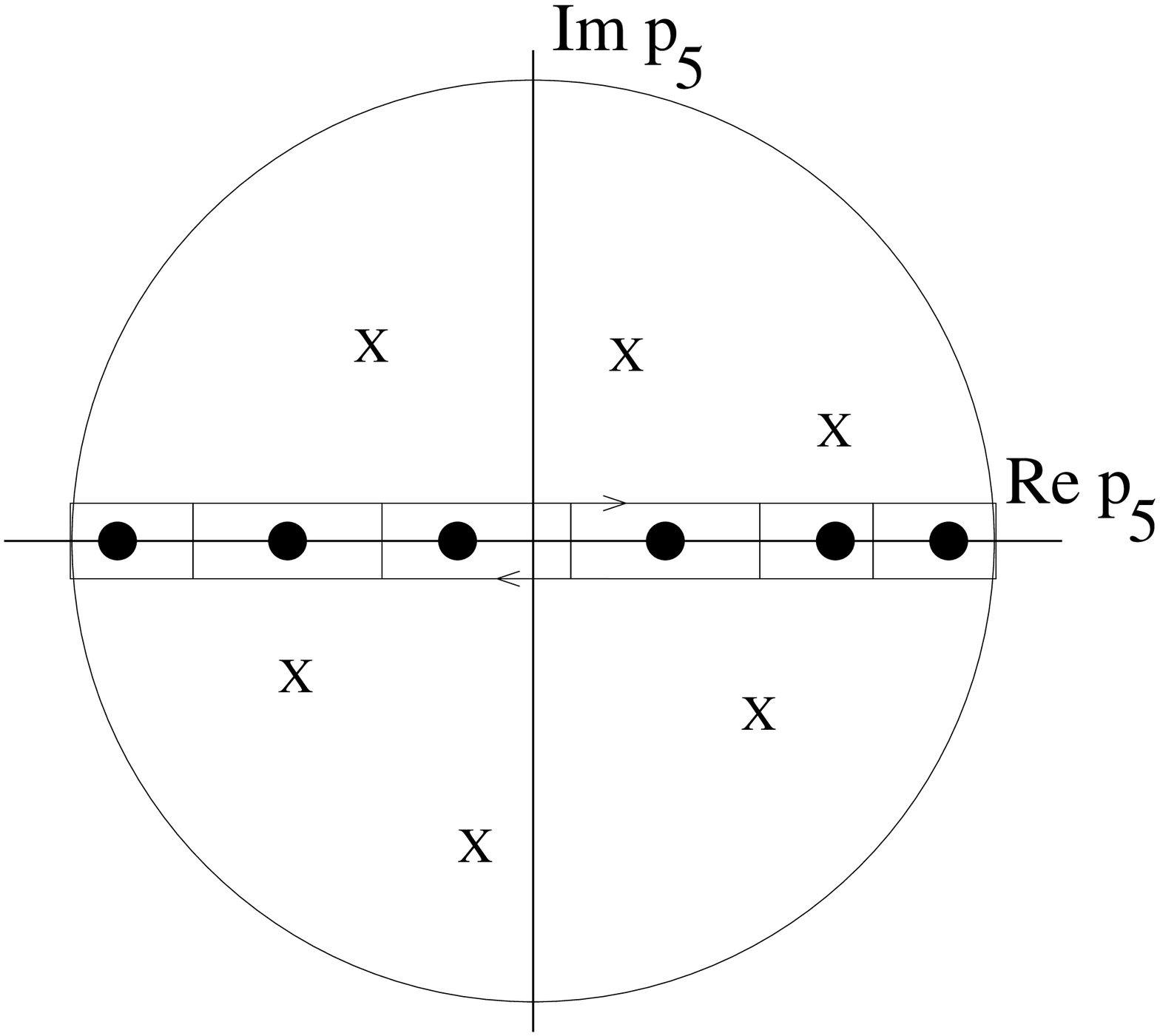, width=4cm, angle =0}
\end{center}
The dots $\bullet$ denote the KK masses $2n/R$; the poles of
$\cP^{++}$ and the $X$'s denote generic poles in the integrant: 
in this case at $\pm i \sqrt{p_4^2 + m^2}$. The regulated
sum-integral is now defined by 
\equ{
\int_\ominus \d^{D_5} p_5 \int \d^{D_4} p_4\, 
 \frac { \cP^{++}(p_5) }{p_4^2 + p_5^2 + m^2} 
\equiv  
 \int_{\ominus} \frac{\d p_5}{2\gp i}\int_0^\infty \d p_4 
\, \cR_{4}( p_4) \cR_{5}(p_5) \, 
\frac{\cP^{++}(p_5)}{p_4^2 +p_5^2+m^2},
}
with complex dimensions $D_4$ and $D_5$ that act as regulators. 
The regulator functions $\cR_{4}(p_4)$ and $\cR_5(p_5)$ are given by
\equ{
\cR_{4}(p_4) =  
\frac {2 \gp^{\half D_4}}{\gG(\half D_4)}\, p_4^3 \, 
\Bigl( \frac {p_4}{\gm_4} \Bigr)^{D_4 -4}, 
\qquad 
\cR_{5}(p_5) =  
\frac {\gp^{\half D_5}}{\gG(\half D_5)}\, 
\Bigl( \frac {p_5}{\gm_5} \Bigr)^{D_5 -1}.
\labl{regufunctions}
}
Here we have introduced two (arbitrary) renormalization scales 
$\gm_4$ and $\gm_5$. 
The regulator function $\cR_{4}$ is the standard one for 
dimensional regularization of 4 non-compact dimensions \cite{tHooft}. 

One of the crucial properties the regularization of the sum-integral
in this way is that the classification of different types of
divergences is independent of the regulator, but depends only on the
spectrum of the KK tower encoded in the pole functions $\cP^{\pm\pm}$. 
In fact, it can be shown that (for ${ \Im\, p_5 > 0}$) the
pole functions consist of three parts \cite{sgn}: 
\equ{
\mtrx{{\cP^{\pm \pm} (p_5)} = & \!\!
{- \frac i4 \gp R} & \!\! {\pm \half {\dsp \frac 1{p_5}}} &
  \!\! - \half \gr_{>}(p_5) 
\\ 
&{\uparrow}& {\uparrow}& {\uparrow}
\\
&{\text{5D}} & {\text{4D}} & {\text{Finite}}
}
}
By inserting only the first part into the regulated sum-integral
expression a cubic divergence arises as one would expect for an
integral over this propagator in 5 dimensions. The second term
represents just a single pole, hence there is no need for a separate
regulator $D_5$ for the sum. Therefore we can safely put it to $1$,
and obtain a quadratic divergence as in 4 dimensions. The last term
gives a finite contribution, because it can be shown that 
$\gr_{>}$ exponentially
suppressed for large complex momenta with $\Im\ p_5 > 0$.  
(For $\Im\ p_5 < 0$ a similar result can be derived.)

\section{The Fayet-Iliopoulos term}

In a (unbroken) supersymmetric field theory in 
4 dimensions the FI-term is either quadraticly divergent or vanishes 
at one loop.  The diagram of the FI-contribution to the selfenergy of a 
scalar is given by:
\begin{center}
\begin{picture}(0,0)%
\includegraphics{FI.pstex}%
\end{picture}%
\setlength{\unitlength}{2763sp}%
\begingroup\makeatletter\ifx\SetFigFont\undefined%
\gdef\SetFigFont#1#2#3#4#5{%
  \reset@font\fontsize{#1}{#2pt}%
  \fontfamily{#3}\fontseries{#4}\fontshape{#5}%
  \selectfont}%
\fi\endgroup%
\begin{picture}(3024,1519)(2089,-5773)
\put(4126,-4561){\makebox(0,0)[lb]{\smash{\SetFigFont{11}{13.2}{\familydefault}{\mddefault}{\updefault}
$h$
}}}
\put(3226,-5536){\makebox(0,0)[lb]{\smash{\SetFigFont{11}{13.2}{\familydefault}{\mddefault}{\updefault}
$D^\parallel$
}}}
\end{picture}

\end{center}
The dotted line corresponds to the auxiliary field $D^\parallel$ 
of the Abelian gauge multiplet in 4 dimensions. In \cite{SHPD} we have
investigated what happens to the FI-term in the effective field theory 
coming from 5 dimensions with a mass spectrum of the complex scalars 
of the hyper multiplet on $S^1/\Intr_2 \times \Intr_2'$. 
We take the charges of these scalars 
such that $q^{++}_n = - q^{--}_n = 1$. Formally, the expression for 
the one loop contribution to the FI term reads 
\equ{
\gx^{} = \sum_{n, \ga} \,g q^{\ga\ga}_n \int \frac{\d^4 p_4}{(2\gp)^4} 
\frac {1}{p_4^2 + ({m_n^{\ga\ga}})^2 + m^2},
\labl{FIoneloopKK}
} 
where $m_n^{\ga\ga} = 2n/R$ and the sum for $\ga = +$ is over 
$n \geq 0$, while for $\ga = -$ over $n > 0$. 
Using dimensional regularization we obtain
\equ{
\gx^{} = g\,  \int \frac{\d^{D_4} p_4}{(2\gp)^{D_4}} 
\int_\ominus \frac { \d^{D_5} p_5}{2\gp i}
\left\{
\frac{\cP^{++}(p_5)}{p_4^2 + p_5^2 + m^2} - 
\frac{\cP^{--}(p_5)}{p_4^2 + p_5^2 + m^2}
\right\}.
}
Substituting the expressions of the pole functions 
\eqref{polefunS22}, gives exactly the same result 
as the regulated FI term for one massless complex scalar:
\equ{
\gx^{} = g\,   \int \frac{\d^{D_4} p_4}{(2\gp)^{D_4}} 
\int_\ominus  \frac {\d^{D_5} p_5}{2\gp i} 
\frac 1{p_5} 
\frac {1}{p_4^2 + p_5^2 + m^2} 
= g\, 
\int \frac{\d^{D_4} p_4}{(2\gp)^{D_4}} \frac {1}{p_4^2 + m^2}.  
\labl{FItermzero}
}
Since it behaves as a single particle contribution we can safely take 
$D_5 =1$ giving the 4 dimensional quadratically divergent expression. 
This result holds for any finite $R$, since it is independent 
of the radius $R$ of the compact dimension. Therefore, we conclude 
that it is also true in the limit $R \ra \infty$. This signals that the 
orbifolding is not undone in this decompactification limit. 

One may wonder whether this divergence may be canceled by other gauge
corrections. In \cite{SHPD} it is shown that the other gauge
contributions give a finite correction and can therefore
never cancel this quadratic divergence. Heuristicly, this is to be
expected since the FI-term is the only diagram of all gauge correction
to the selfenergy that is proportional to the trace of the charges in
the loop. 

For this Fayet-Iliopoulos contribution, an auxiliary field tadpole counter 
term has to be introduced. 
Such a counter term of course has to be consistent with the symmetries 
of the theory. On both branes we have at most $N = 1$ supersymmetry:  
the other supersymmetry vanishes there as it has the opposite parity.
Therefore, on the branes $D$-terms can be added for the 
auxiliary fields that do not vanish. By a similar analysis on the
effective field theory level one can indeed show that one obtains
the same quadratic divergence for all $D^\parallel_{2n}$. In the 5
dimensional picture this means that the divergences occur on the two
branes only.

\section{Low energy anomalies} 

Gauge anomalies render a theory to be
inconsistent at the quantum level. Also in models with one extra
dimension that are under investigation here, one has to address the
question of low energy anomalies. Therefore in the effective
field theory in 4 dimensions the triangle diagram 
\[
{\sum_\gps} 
\raisebox{-8mm}{$\begin{picture}(0,0)%
\includegraphics{triangle.pstex}%
\end{picture}%
\setlength{\unitlength}{1184sp}%
\begingroup\makeatletter\ifx\SetFigFont\undefined%
\gdef\SetFigFont#1#2#3#4#5{%
  \reset@font\fontsize{#1}{#2pt}%
  \fontfamily{#3}\fontseries{#4}\fontshape{#5}%
  \selectfont}%
\fi\endgroup%
\begin{picture}(5424,2433)(4489,-5921)
\put(5251,-4936){\makebox(0,0)[lb]{\smash{\SetFigFont{6}{7.2}{\familydefault}{\mddefault}{\updefault}
$A_\gm$
}}}
\put(8851,-4036){\makebox(0,0)[lb]{\smash{\SetFigFont{6}{7.2}{\familydefault}{\mddefault}{\updefault}
$A_\gn$
}}}
\put(8851,-5836){\makebox(0,0)[lb]{\smash{\SetFigFont{6}{7.2}{\familydefault}{\mddefault}{\updefault}
$A_\gr$
}}}
\put(7126,-4786){\makebox(0,0)[lb]{\smash{\SetFigFont{6}{7.2}{\familydefault}{\mddefault}{\updefault}
$\gps$
}}}
\end{picture}
$}
\]
has to be considered with the sum over all (chiral) fermion
spices. This includes summing over all KK excitations since of course
in the loop heavy virtual particles may run around. 

The standard way of thinking about anomalies is that only the
zero-mode fermions can contribute. Hence, if the massless fermionic
spectrum is anomaly free, then no anomalies in the low-energy field
theory can arise. In particular, the BHN model having the fermionic
spectrum of the standard model is anomaly free. 
Using the dimensional regularization procedure of a compact dimension
discussed above this naive expectation is indeed confirmed that there
are no gauge anomalies in this model. 

These anomaly are on the level of the low-energy effective field theory. 
In ref.\ \cite{Scrucca:2001eb} a more subtle form of anomaly is 
discussed which is localized at the both fixed points where
opposite chiral states are projected out, while the integrated anomaly
vanishes. However, it still has to be clarified how these fixed point
anomalies exactly effect the low-energy physics.

\section{Conclusion}

In this talk we have discussed an extension of dimensional
regularization that can be applied to a (factorizable) space-time that
has one compact dimension,  using complex contour integral to represent
(divergent) sums. This contour integral can then be regularized by
introducing a regulator function inspired by standard dimensional
regularization. Having two regulators ($D_4$, $D_5$), this 
regularization prescription treats the additional dimension 
without any prejudice. 
And in addition it leave the properties of the KK towers in tact, 
since they are encoded in regularization independent pole functions. 

Using this method we showed that the tadpole contribution to a 
component of the auxiliary field is quadratically divergent and 
proportional to the sum of (hyper) charges 
of massless scalar fields in the BHN model. 
It is not difficult to identify the 5 and 
4 dimensional divergent and finite contribution, exploiting 
the properties of the pole functions 
associated with the orbifold. 
The behavior of the FI-tadpole is very similar to the low-energy
anomalies in the sense that they are both independent of the
size of the extra dimension. In particular, even if we take
$R\ra\infty$ we find the same expression due to the massless modes.

\section*{Acknowledgements} 

The author would like to thank H.P.\ Nilles and D.\ Ghilencea for the
stimulating collaboration during part of this work. 
This work is supported by priority grant 1096 of the Deutsche 
Forschungsgemeinschaft and European Commission RTN 
programmes HPRN-CT-2000-00131 / 00148 and 00152.

\end{document}